\documentclass[runningheads]{llncs}
\usepackage{subfigure}
\usepackage{graphicx}
\usepackage[T1]{fontenc}
\usepackage{graphicx}
\begin{document}
\title{Uncertainty, bias and the institution bootstrapping problem}
%
%
\author{Stavros Anagnou\inst{1,2} \and
Christoph Salge\inst{1} \and
Peter R. Lewis\inst{2}}
\authorrunning{S. Anagnou et al.}
%
\institute{Adaptive Systems Research Group, University of Hertfordshire, United Kingdom  \and
Trustworthy AI lab, Ontario Tech University, Canada
\\\email{s.anagnou@herts.ac.uk}\\
}
\maketitle              
\begin{abstract}
Institutions play a critical role in enabling communities to manage common-pool resources and avert tragedies of the commons. Prior research suggests institutions emerge when universal participation yields greater collective benefits than non-cooperation. However, a fundamental issue arises: individuals typically perceive participation as advantageous only after an institution is established, creating a paradox—how can institutions form if no one will join before a critical mass exists? We term this conundrum the institution bootstrapping problem and propose that misperception—specifically, agents’ erroneous belief that an institution already exists—could resolve this paradox. By integrating well-documented psychological phenomena—including cognitive biases and perceptual noise—into a game-theoretic framework, we demonstrate how these factors collectively mitigate the bootstrapping problem. Notably, unbiased perceptual noise (e.g., noise arising from agents’ differing heterogeneous physical or social contexts) drastically reduces the critical mass of cooperators required for institutional emergence. This effect intensifies with greater diversity of perceptions, suggesting that variability among agents perceptions facilitates collective action. We explain this counter-intuitive result through asymmetric boundary conditions: proportional underestimation of low-probability sanctions produces distinct outcomes compared to equivalent overestimation. Furthermore, the type of perceptual distortion—proportional versus absolute yields qualitatively different evolutionary pathways. These findings challenge conventional assumptions about rationality in institutional design, highlighting how "noisy" cognition can paradoxically enhance cooperation. Our analysis highlights how, even though biases and uncertainty are often perceived as defects of human cognition, they create the cognitive conditions that enable institutions to bootstrap into existence.

\keywords{Uncertainty \and Bias \and Institutions \and Bootstrap \and Evolutionary Game Theory \and Noise \and Bounded Rationality }
\end{abstract}

\clearpage
\section{Introduction}
\subsection{The institution bootstrapping problem}
Across human societies, it is difficult to find groups without some form of institution \cite{north_institutions_1990}. Institutions—defined as rule systems enabling sustainable management of common-pool resources (e.g., grazing lands, fisheries, or even Minecraft servers)—help communities avoid antisocial outcomes by regulating individual self-interest \cite{ostrom_governing_1990,frey_emergence_2019}. There are many empirical examples of institutional rules that help groups avoid anti-social outcomes. These rules regulate individual self-interest so that resources can be managed sustainably. For instance, they may set limits on how much water a person can take from a shared irrigation system and how often they must perform maintenance \cite{north_institutions_1990,ostrom_governing_1990}. Critically, institutions are self-reinforcing: both compliance and active enforcement align collective good with individual incentives, as demonstrated in foundational studies \cite{north_institutions_1990,ostrom_governing_1990}.

However, institutions require sustained effort to create, maintain, and adapt. Without this, groups risk reverting to default interactions where cooperation collapses. Ostrom’s fieldwork emphasises that institutions are more likely to endure when rules are both designed and enforced by the same agents whose actions they govern \cite{ostrom_governing_1990}. Institutional roles—such as monitoring compliance or coordinating rule updates—incur costs that must be offset by collective benefits in order for self-interested individuals to join. Recent work formalises these dynamics \cite{pitt_axiomatization_2012,powers_modelling_2018}. For example, Powers et al \cite{powers_modelling_2018} use evolutionary game theory (EGT), to derive conditions where institutional participation becomes evolutionarily stable. For instance, monitoring costs must remain low relative to resource contribution costs, and institutional roles (e.g., monitors) must be incentivised through resource-sharing mechanisms \cite{powers_modelling_2018}.


A critical challenge arises even when institutions are theoretically favourable: the institution bootstrapping problem. As Powers et al \cite{powers_modelling_2018} notes, institutions require a threshold of participants to generate sufficient benefits (e.g., monitoring capacity) to justify individual costs. Without this critical mass, free-riding dominates because early adopters bear disproportionate costs (e.g., monitoring efforts) without guaranteed reciprocity.

So, if we assume all individuals in a group are not yet part of the institution, the benefits do not outweigh the costs of joining. Since there are no monitors to enforce rules on peers, there is a lesser incentive to contribute as opposed to free-riding. We term this issue the institution bootstrapping problem. We consider several explanations for this and possible approaches for alleviating or completely overcoming the problem, where agents would manage to bootstrap their way towards the institution despite the incentives against it.

Proposed solutions include extrinsic shocks (e.g., external cooperator influx) or cognitive factors like cost misperception \cite{powers_modelling_2018}. Small groups might circumvent the problem through charismatic leadership or evolved psychological mechanisms that amplify cooperation, e.g. trust \cite{lewis_what_2022}. Our approach focuses on relaxing perfect rationality assumptions in evolutionary game theory by incorporating perceptual biases and uncertainty in the form of noise inherent to bounded agents operating in heterogeneous physical/social environments.



\subsection{Contributions}
Using an evolutionary game theoretic approach we: 
\begin{enumerate}
    
\item Explicitly illustrate, for the first time, the bootstrapping problem in a simplex plot.
\item Show that incorporating a coarse bias into how agents perceive the cost of freeriding can either decrease or increase the number of cooperators and monitors needed to establish the institution. Overestimating the risk of punishment leads to greater cooperation. This holds if we generally assume that the agents are subject to a loss aversion bias, but given complicated emerging empirical evidence on how individuals perceive probabilities, we need a more solid empirical foundation to motivate this bias.
\item Show that, by incorporating a more nuanced S-shaped or inverse-S-shaped probability distortions from the literature on human psychophysics, we can similarly show that this affects the threshold of cooperators needed to establish an institution. However, limitations due to the individual context of the experiments and task dependency of the effect complicate interpretation. We, therefore, urge such psychophysics experiments to take place in a social context, as their implications would matter for the psychology of enduring institutions.
\item Show that incorporating noisy perception to capture the bounded heterogeneity among agents (e.g. their differing social circles and positions in the physical world) decreases the threshold of cooperators needed to establish the institution. Interestingly, this is despite the noise being unbiased at the group level.
\item  Explain the above counter-intuitive result in terms of an asymmetric boundary condition where proportionally underestimating very small quantities is not the same as overestimating them and show that the type of noisy perception (proportional or absolute), results in different qualitative results. 

\end{enumerate}


\section{Modelling the institution bootstrapping problem}

In this section, we will describe the bootstrapping problem in terms of evolutionary game theory (EGT).

To capture the bootstrapping problem in more concrete terms, we will express it in the form of a game-theoretic model. Investigating such theories in a mathematical framework helps us explicitly define our assumptions and system specifications. It allows us to assess the logic of our ideas and establish whether they are internally coherent enough to serve as a good candidate explanation. This added rigour helps us avoid logical errors or missed details due to the inherent ambiguity of verbal theorising.

Further, the abstractions in a mathematical model allow us to capture dynamics common across many types of institutions and provide a general understanding of how they work without getting lost in details. This gives us a solid theoretical foundation to subsequently incorporate the specifics of each situation.

We adapt the Powers model of institutions \cite{powers_modelling_2018}, using it to describe explicit utility functions within an evolutionary game theory (EGT) framework for each action \cite{smith_evolution_1982}. EGT evaluates the utility of an agent’s actions, which reflects the material or psychological consequences of those actions. These utility functions are then combined with a replicator equation to plot the rate of change in the relative frequencies of strategies, generating a simplex plot that visualises population dynamics across states (i.e., the number of agents adopting each strategy).

We outline three strategies available to individuals: defector (D), contributor (C), and contributor-monitor (CM). A \textbf{defector (D)} does not engage in the institution, undermining it by consuming from the common-pool resource without contributing (\(C_c\)) and avoiding institutional roles. Defectors gain benefits from the common pool (\(B_g\), Table~1) but incur a free-riding cost (\(C_f\), Table~1) in the form of punishment from monitors.
This cost scales with \(N_m/N\), the fraction of monitors in the population, reflecting the increased likelihood of being caught as the proportion of monitors rises (Table~2). It also scales with \(p\), the number of checks a monitor makes and \(s\), the cost of punishment. 
This weighted expectation of being punished (\(C_f\), as captured abstractly by a reduction in utility, models the potential consequences of peer punishment, e.g. a material fine, the feeling of shame or damage to reputation) \cite{savarimuthu_norm_2011,nardin_classifying_2016}. This deters individuals from defecting and maintains social order \cite{nardin_classifying_2016}.

The \textbf{contributor (C)} participates in the institution, benefiting from it (\(B_g\), Table~2) while paying a contribution cost (\(C_c\), Table~2) determined by the parameter \(\alpha\), which quantifies the individual’s contribution level (Table~2).  

The \textbf{contributor-monitor (CM)} also participates and contributes, but additionally takes on a monitoring role to enforce institutional rules. Like contributors, CM agents benefit (\(B_g\)) and pay contribution costs (\(C_c\)). Monitoring incurs an additional cost (\(C_m\), Table~2), which includes effort (e.g., time spent checking compliance) or risks (e.g., retaliation from punished defectors). These costs are modelled as \(p \cdot \delta\), where \(p\) represents the number of monitoring checks performed, and \(\delta\) is the cost per punishment instance. Monitors receive a benefit (\(B_m\), Table~2) proportional to \(\beta\) (the share of common-pool resources allocated to monitors), \(\alpha\) (individual contributions), and \(N_c/N_m\) (the contributor-to-monitor ratio), with larger \(N_m\) diluting individual shares.

\begin{table}[]
\caption{Table outlining strategies in terms of their role, real-world example and utility payoff in the model }
\label{tab:my-table}
\resizebox{\columnwidth}{!}{
\begin{tabular}{|l|l|c|c|}
\hline
Strategy & Equation                       & \multicolumn{1}{l|}{Role}                                                                                                                 & \multicolumn{1}{l|}{Real world examples (local town)}                                                                                                        \\ \hline
D        & $U_D = B_g - C_f$              & \begin{tabular}[c]{@{}c@{}}Agent who is depleting common pool resource\\ but not contributing to it, is punished by monitors\end{tabular} & \begin{tabular}[c]{@{}c@{}}Individual uses public infrastructure \\ but doesn't pay to upkeep it\end{tabular}                                                \\ \hline
C        & $U_C = B_g - C_c$              & \begin{tabular}[c]{@{}c@{}}Agent contributing to common pool resource \\ but not taking on a monitoring role to punish D\end{tabular}     & \begin{tabular}[c]{@{}c@{}}Individual uses public infrastructure \\ and pays to upkeep it\end{tabular}                                                       \\ \hline
CM       & $U_{CM} = B_g - C_c + B_m - B_c$ & \begin{tabular}[c]{@{}c@{}}Agent contributing to common pool resource \\ and taking on a monitoring role to punish D\end{tabular}         & \begin{tabular}[c]{@{}c@{}}Individual uses public infrastructure, \\ pays to upkeep it \\ and goes out of their way to punish others who do not\end{tabular} \\ \hline
\end{tabular}}
\end{table}

\begin{table}[]
\caption{Table describing the associated costs and benefits of each action along with an equation defining it.}
\label{tab:my-table}
\resizebox{\columnwidth}{!}{
\begin{tabular}{|l|l|l|}
\hline
Cost or Benefit & Meaning                                                                                                                      & Equation                         \\ \hline
$B_g$           & \begin{tabular}[c]{@{}l@{}}Benefit from collective \\ resource of being a member of the group\end{tabular}                   & $B_g = (1-\beta) \cdot 1/N\cdot\alpha \cdot N_c$ \\ \hline
$C_c$           & Cost of contributing                                                                                                         & $C_c = \alpha$                   \\ \hline
$B_m$           & Benefit of monitoring                                                                                                         & $B_m = \alpha\cdot\beta\cdot N_c / Nm$    \\ \hline
$C_m$           & Cost of monitoring                                                                                                           & $C_m = p \cdot \delta$               \\ \hline
$C_f$           & \begin{tabular}[c]{@{}l@{}}Cost of free-riding (note that this a expected cost \\ conditioned on a probability)\end{tabular} & $C_f = (p\cdot N_m/N )\cdot s$             \\ \hline
\end{tabular}}
\end{table}

\begin{table}[]
\caption{Table describing each world parameter that is used to compute the utility of each strategy}
\label{tab:my-table}
\resizebox{\columnwidth}{!}{
\begin{tabular}{|l|l|l|l|}
\hline
Parameter             & Meaning                                            & \begin{tabular}[c]{@{}l@{}}Value (to favour \\ institution)\end{tabular} & \begin{tabular}[c]{@{}l@{}}Value (to not favour \\ institution)\end{tabular} \\ \hline
$\alpha$ & Cost of contributing to common pool resource       & 1                                                                        & 1                                                                            \\ \hline
$\beta$  & Fraction of common pool resource given to monitors & 0.2                                                                      & 0.2                                                                          \\ \hline
$\delta$& Cost of punishment to monitor                      & 0.1                                                                      & 0.5                                                                          \\ \hline
$p$                   & Number of checks each monitor makes                & 5                                                                        & 5                                                                            \\ \hline
$s$                   & Cost of being punished                             & 1                                                                        & 1                                                                            \\ \hline
$N_m$                 & Number of monitors                                 & Dynamic                                                                  & Dynamic                                                                      \\ \hline
$N_c$                 & Number of contributors (includes CMs)               & Dynamic                                                                  & Dynamic                                                                      \\ \hline
$N$                   & Total number of agents                             & 20                                                                       & 20                                                                           \\ \hline
\end{tabular}}
\end{table}

\begin{figure}[hbt!]
    \centering
    \subfigure[a]{\includegraphics[width=0.29\textwidth]{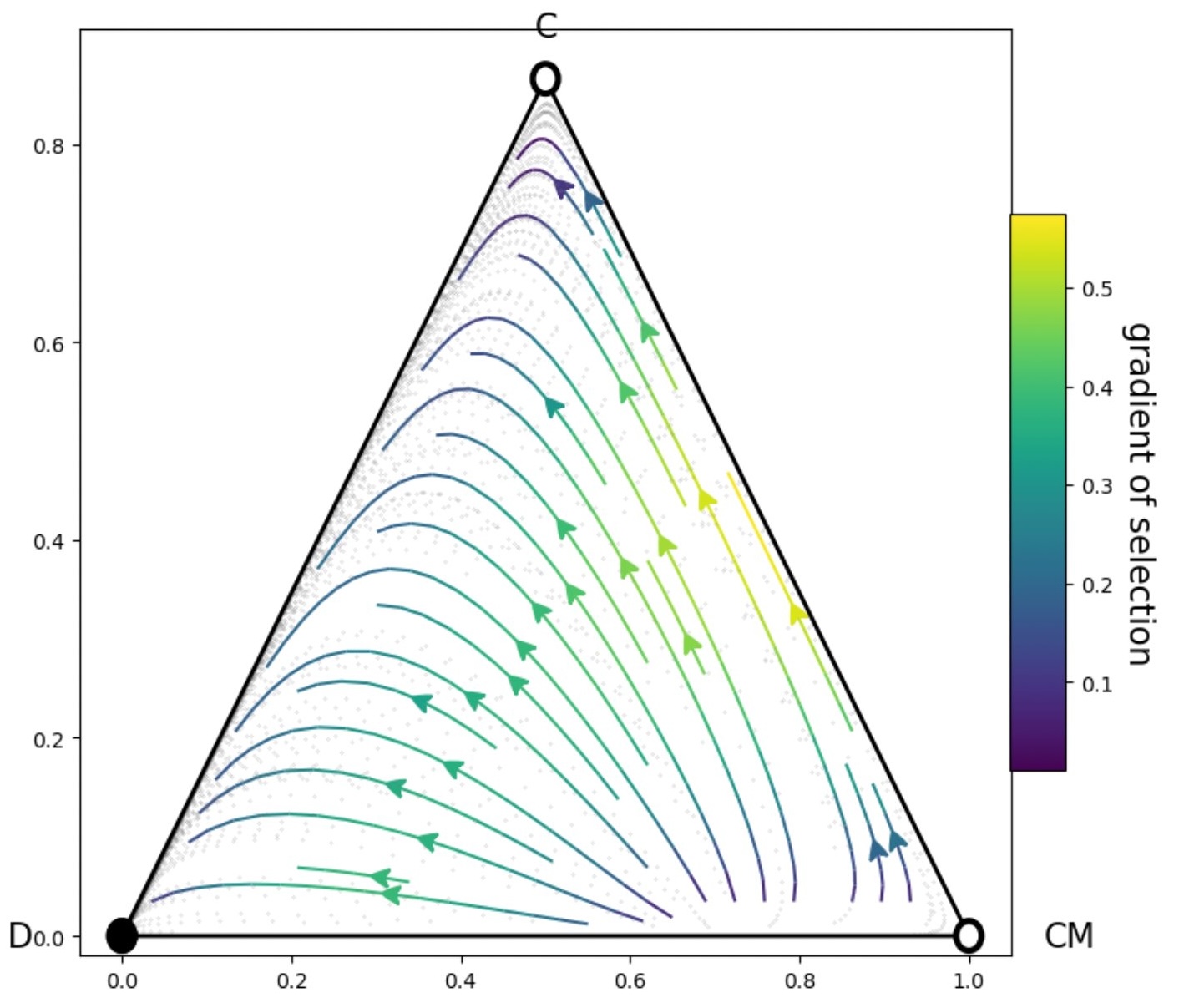}} 
    \subfigure[b]{\includegraphics[width=0.29\textwidth]{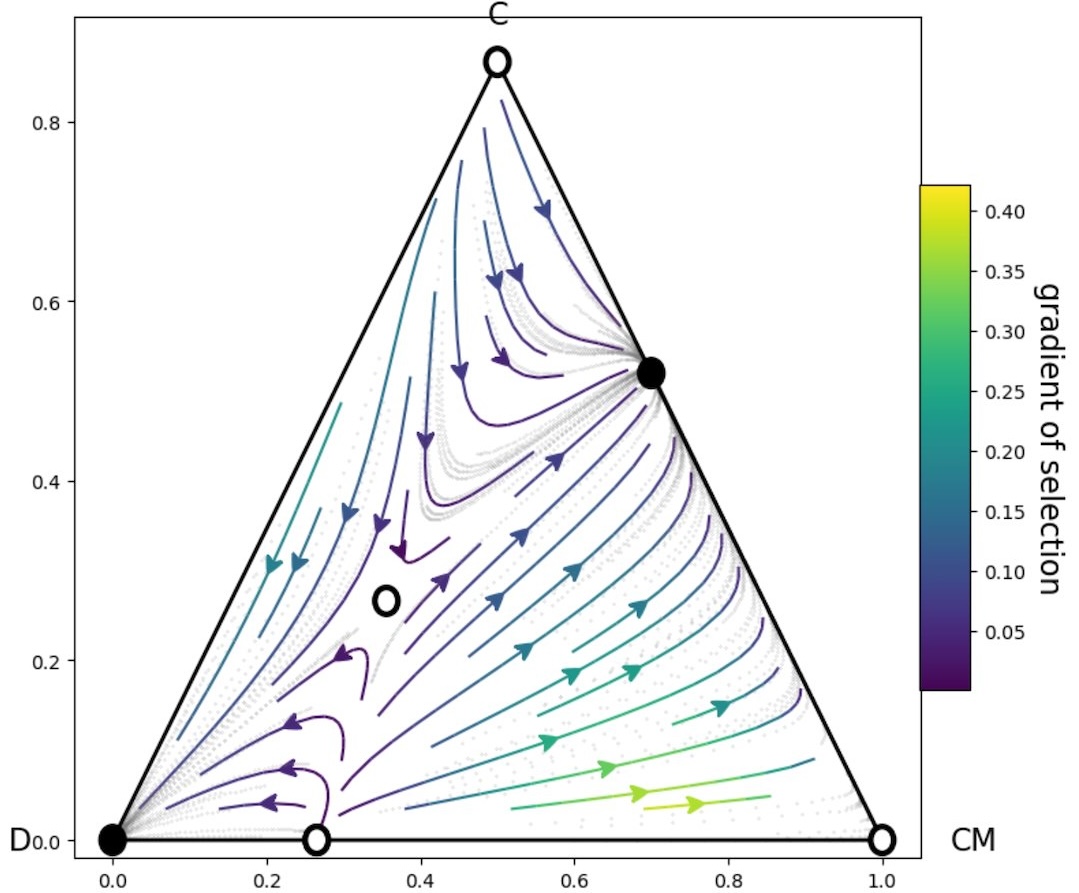}} 
    \caption{(a) depicts the simplex plot for 3 strategies where the inequality specified by Powers et al  \cite{powers_modelling_2018} (a mathematical inequality dependent on world parameters that determines whether or not an institution will be formed) is not satisfied and an institution does not form (b) depicts a case where the inequality from Powers \cite{powers_modelling_2018} where an institution does form, however for areas in strategy space closer to all defectors, the system tends toward all defectors. This means that although conditions in terms of parameters are favourable for an institution to form, a critical mass of individuals joining the institution is needed before it will form. Reaching this threshold is the institution's bootstrapping problem. The black circles depict unstable attractors, and the black dots signify stable attractors 
    }
    \label{fig1}
\end{figure}

We have described a system where the consequences of an agent's actions depend on facts in the world, e.g. the cost of punishment $s$ and the frequency of other strategies in the population, e.g. the number of monitors determines how much an agent is punished (Tables 1,2,3). This means the utility of each individual depends on what other individuals are doing.

The utilities of each strategy (D,C,CM) determine the strategy's success and their propagation through the population. In classic EGT this is cast in terms of genetic evolutionary fitness. But we can also interpret this to be cultural fitness, i.e. strategies that are better off would be more likely to be copied by others through prestige-biased social learning \cite{boyd_culture_1988}. We implement this using the replicator equation (equation 1).

The replicator equation computes the gradient for a given group configuration $x_i$. It does this by multiplying the average fitness of a strategy across at the population at a given group configuration  $\Pi_i(\mathbf{x})$ vs the average fitness of all strategies for that given group configuration $\overline{\Pi}(\mathbf{x})$ (equation 1).

\begin{equation} \label{eu_eqn}
\dot{\mathbf{x}}_i = \mathbf{x}_i \left( \Pi_i(\mathbf{x}) - \overline{\Pi}(\mathbf{x}) \right)
\end{equation}
We can use the replicator equations to plot the direction in strategy space for each group combination in a simplex and show when agents will tend to the pro-institution strategies (C and CM) and then they will not (D). We do this using the library EGTtools \cite{fernandez_domingos_egttools_2023}.

Powers derived conditions under which an institution will endure, expressed through mathematical inequalities \cite{powers_modelling_2018}. Specifically, these inequalities compare institutional benefits against participation costs: an institution endures when benefits exceed costs. 

In Fig. 1, we plot simplexes representing strategy space gradients for systems that do and do not satisfy these inequalities, illustrating the bootstrapping problem. Arrows indicate selection gradients, showing the direction of strategy propagation under evolutionary dynamics. The corners (D, C, CM) correspond to homogeneous populations (all agents adopting one strategy), while the centre represents equal strategy proportions. White dots denote unstable equilibria (system states persisting only without stochastic changes, e.g., imitation errors during strategy updates). Black dots signify evolutionary stable strategies (ESS), robust to perturbations in group composition and therefore cannot be invaded by another strategy. 

In Fig. 1a, the simplex corresponds to parameters predicted to preclude institutional formation per \cite{powers_modelling_2018}. Here, institutions fail to form in almost all configurations, with no stable attractor for populations of C or CM. The sole stable attractor occurs at full defection (D), indicating that an institution will not form in almost any case, and, if it does, it will be fragile to invasion by non-cooperative strategies.

Fig. 1b shows parameters predicted to enable institutional formation per \cite{powers_modelling_2018}. A large region of the strategy space converges toward an ESS for institutional cooperation (C/CM dominance), where defection yields no advantage. This ESS resists large perturbations in group composition. However, a smaller region persists where insufficient monitors/cooperators render cooperation non-beneficial, resulting in a D-dominated ESS. In this region, isolated strategy shifts (e.g., one agent cooperating) revert to D due to a lack of collective incentives.  

Thus, we have captured the institution bootstrapping problem: despite the institution being beneficial to join, if the system starts in an initial non-cooperative situation (which is an equilibrium), agents need a catalyst to be able to bootstrap the creation of institutions, which can then lead to socially preferable, sustainable outcomes (also shown to be an equilibrium).

\section{Incorporating perception into social simulation, an unlikely solution to the bootstrapping problem?}

How do we get around the bootstrapping problem? One can assume an influx of cooperators/monitors to the group which then incentivises others to join, or a strong leader type emerges which forces the requisite number of individuals As seen in the equation for the expected cost of free riding $C_f$ (Table 2), it assumes a perfect perception of what would often be hidden variables, e.g. the number of checks a monitor makes $p$ or the number of monitors in the population at any given time. approach: Can agents merely pretending an institution exists, make it a reality?

We will now try to alleviate the bootstrapping problem by questioning the cognitive assumptions of EGT.

As seen in the equation for the expected cost of free riding $C_f$ (Table 2), it assumes a perfect perception of what would often be hidden variables, e.g. the number of checks a monitor makes $p$ or the number of monitors in the population at any given time.

It is very difficult for bounded agents with noisy perception to have a perfect estimate; in most cases, they often resort to heuristics in the form of biases \cite{tversky_advances_1992,simon_behavioral_1955,hertwig_fast_2009}.
Emerging work in game theory and social simulation \cite{mcnamara_game_2022,vogrin_confirmation_2023} shows that bringing even simple psychological facets in can change the predictions of game theory and the dynamics of agent-based models \cite{mcnamara_game_2022}.

We focus on bias in the literature of social modelling and evolutionary game theory.

Firstly, there is work showing that bias, even though it may be misrepresentative of reality, may be advantageous to individuals. For example, in a frequency-dependent hawk-dove game, where the prevalence of doves and hawks influences strategic payoffs, overestimating the reward of a hawk action can mislead unbiased individuals into perceiving their own actions as less advantageous. As a result, they alter their behaviour, indirectly benefiting biased hawk-strategy players, who now face fewer competitors \cite{mcnamara_learning_2021}. Similarly, in resource competition, individuals who are overconfident in their ability to compete are evolutionarily stable across a wide range of environments, particularly under uncertainty \cite{johnson_evolution_2011}.

In the above cases, even though biased individuals were better off, bias was detrimental for group utility, but there are also cases where it is beneficial for the group as well. For example, Vogrin et al \cite{vogrin_confirmation_2023} show that having a bias enhances performance in a signal detection task and suggest that in multiple agent settings, different agents can do a cognitive division of labour by specialising in different signals, thus promoting discourse and a broader investigation of problem spaces. Therefore being beneficial to the whole group. Davies et al \cite{davies_if_2011} show that incorporating an adaptive bias in how one's utility is perceived in a coordination problem can change the attractor landscape of a system, widening the basin of the global optimum and therefore making it more likely for collective systems to arrive at the global optima and therefore benefit all. 
We hope to apply similar tactics to enlarge the cooperative attractor/reduce the number of individuals needed to set up an institution in our scenario.

\section{Experiments}

We modify the expected cost of freeriding $C_f$ in various ways by biasing the value away from the perfect expected probability to model cognitive biases and aspects of an agent’s perceptual uncertainty (noise) and boundedness.

\begin{enumerate}
    
\item Coarse-grained bias: bias perception of $C_f$ (expected cost of freeriding) by multiplying by 0.75 or 1.5 (for under or over-estimating) 
\item Distorting perception of extreme probabilities: The inverse S and S shaped probability distortion curves are modelled by the Prelec function, given by  $\pi(p) = e^{-\zeta (-\ln p)^\lambda}$, where $\pi(p)$ represents the subjective probability, and $p$ is the objective probability within the range $(0 < p \leq 1)$. The parameter $\lambda$ controls the curvature of the function: if $\lambda < 1$, the function follows an inverted-S shape (we used 0.8), meaning small probabilities are overweighted while large probabilities are underweighted; if $\lambda > 1$, (we used 1.2) the function follows an S-shape, where small probabilities are underweighted and large probabilities are overweighted. The parameter $\zeta$ adjusts the elevation of the curve, typically set to 1 in standard applications.
\item  Noise (proportional), here we sample equally above >1 and <1 when doing the fitness calculations in order to simulate a noisy perception as would occur with agents bounded by individual and social context. $noise \sim \mathcal{U}(a, b)$ which is then multiplied by $C_f$
\item   Noise (absolute): Here we do the same as 3, but we add the noise so it is absolute (we add or subtract in the given range).

\end{enumerate}

\subsection{Moran process for simulating finite populations}
To model these effects of noisy bias, we need a stochastic payoff for the defect strategy, which is not possible in a replicator equation, as they assume infinite populations and are deterministic. 

Therefore, we will be using a Moran process, which assumes finite populations and therefore allows for stochastic effects.

We use the library EGTtools to implement the Moran process \cite{fernandez_domingos_egttools_2023}. For smaller population sizes, stochastic effects dominate, necessitating discrete birth-death processes to model behavioural dynamics. This framework introduces the \textbf{finite population selection gradient} \( G(k/Z) \), defined as the difference between probabilities of incrementing or decrementing a strategy’s count, stochasticity intensifies with behavioural ``mutations'' (e.g., imitation errors).  

We model social learning dynamics via a stochastic birth-death process paired with a pairwise comparison rule. At each timestep, a randomly selected individual \( j \) (strategy \( j \)) revises their strategy by potentially imitating a randomly chosen individual \( i \). Imitation probability \( p \) follows the \textbf{Fermi function}:  
\begin{equation} 
p = \frac{1}{1 + e^{\gamma \left( f_i(k_i) - f_j(k_j) \right)}}  
\end{equation}
where \( f_i(k_i) \) and \( f_j(k_j) \) denote the fitness of individuals \( i \) and \( j \), dependent on their strategy abundances \( k_i \) and \( k_j \). Due to finite populations, absolute counts \( k_i \) replace frequencies \( x_i \), where \( x_i \equiv k_i/Z \).  

Here, \( \gamma \) (inverse temperature) modulates selection intensity and imitation accuracy: \( \gamma \to 0 \) induces near-random drift; \( \gamma \to \infty \) renders imitation deterministic. A mutation rate \( \mu \) allows random strategy exploration. Collectively, this adaptive process forms a Markov chain with state transitions governed by strategy fitness and abundance. For specifics, see \cite{fernandez_domingos_egttools_2023}.

\section{Results} 
\subsection{Coarse bias}

Due to limited cognitive resources, agents usually rely on imperfect but frugal heuristics e.g. loss aversion heuristic \cite{mcnamara_risk-sensitive_1992,tversky_advances_1992}.

We incorporate such a bias into our model as a proportional bias on the $C_f$ and see if this changes the attractor landscape and helps address the bootstrapping problem.

\begin{figure}[hbt!]
    \centering
    \subfigure[a]{\includegraphics[width=0.29\textwidth]{base_control_replicator.jpeg}} 
    \subfigure[b]{\includegraphics[width=0.29\textwidth]{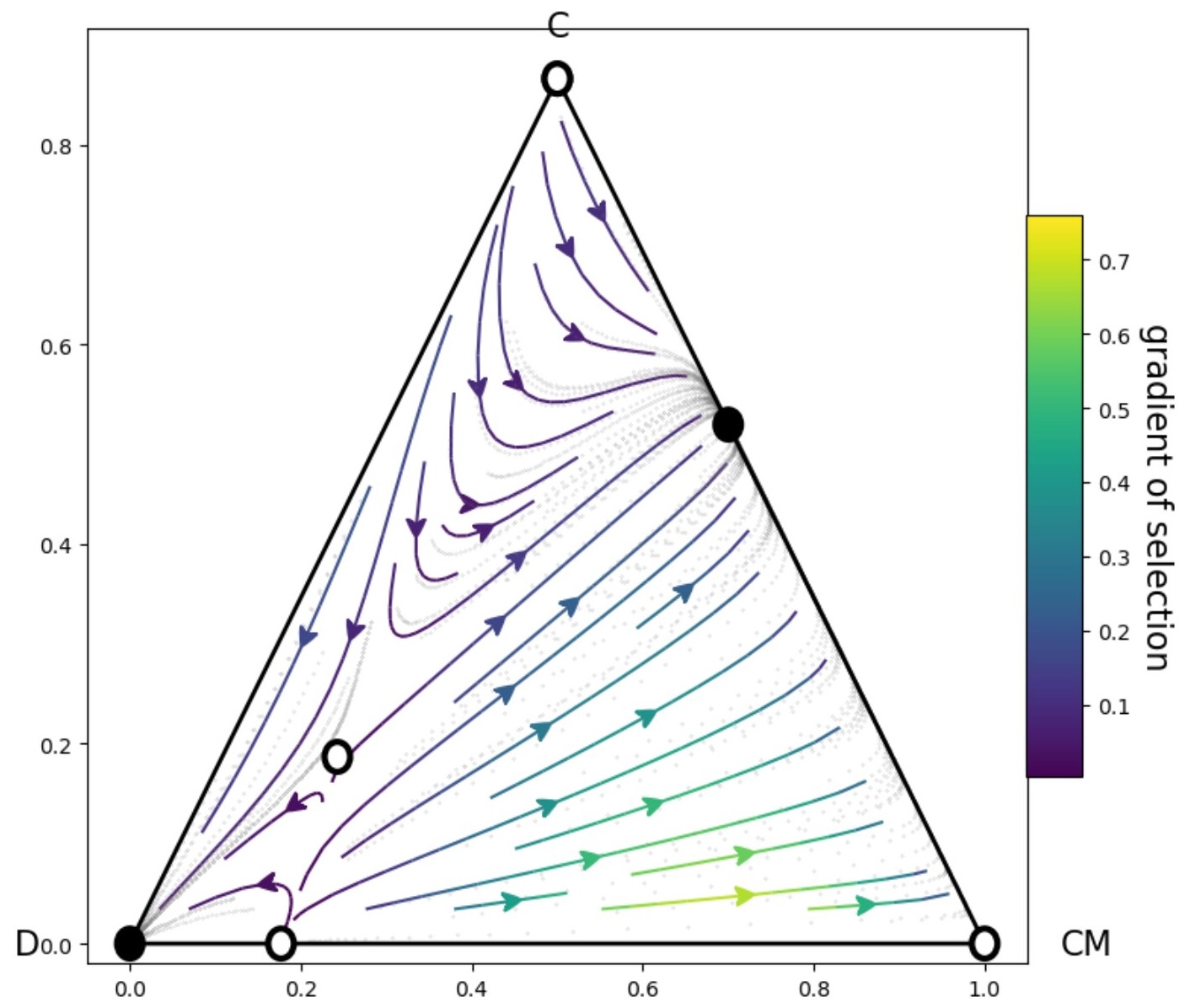}} 
    \subfigure[c]{\includegraphics[width=0.29\textwidth]{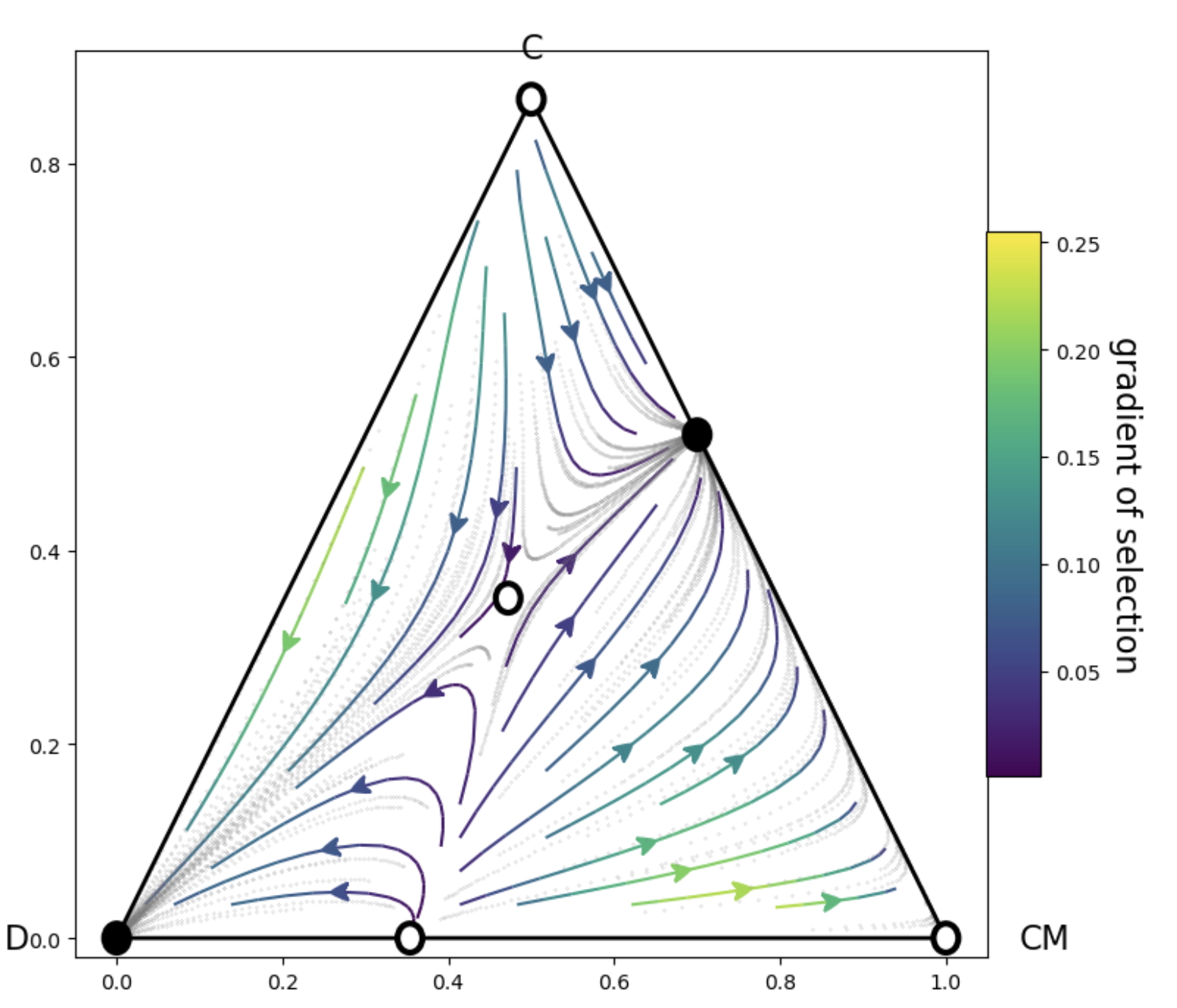}}
    \caption{(a) Control (b) Agents overestimate the expected cost of punishment (c) Agents underestimate the expected cost of punishment}
    \label{fig2}
\end{figure}

We see in Fig.\ref{fig2}b, where agents overestimate the risk of being punished, that the defector attractor indeed shrinks, making it easier for influxes of cooperators or other extrinsic shocks of the system to be able to meet the critical mass of joiners needed to form an institution. For the sake of completeness, we show in Fig.\ref{fig2}c that underestimating the risk leads to a larger defective attractor, and therefore it makes it harder for institutions to form.

 It has been shown empirically that humans overestimate unlikely punishments due to loss aversion bias \cite{mcnamara_risk-sensitive_1992,tversky_advances_1992}, which then would help institutions form. Further, evidence shows that humans overestimate vivid but rare risks, e.g. as can be induced by media sensationalism \cite{sundh_human_2024}. This could explain why public punishment/moral panics are effective ways to influence people to join institutions. We could then argue that this general risk aversion evolved and was then co-opted for institutions. However, more work has been done on bias that complicates this picture of human bias \cite{tversky_advances_1992,zhang_bounded_2020}. We will incorporate this work in the next section.

\subsection{Bias in the form of probability distortion}

 Empirical work has also shown that humans often distort their perception of probabilities at the extremes in a non-linear manner (i.e. distort very small and very large probabilities)  \cite{zhang_bounded_2020}. This results in either an inverted S-shaped distortion (overvalue small probabilities and underestimate large ones) or an S-shaped distortion (undervalue small probabilities and overestimate large ones). We incorporate S and inverse S distortion from the experimental literature into our model to see how it affects the likelihood of institution formation.

\begin{figure}[hbt!]
    \centering
    \subfigure[a]{\includegraphics[width=0.29\textwidth]{base_control_replicator.jpeg}} 
    \subfigure[b]{\includegraphics[width=0.29\textwidth]{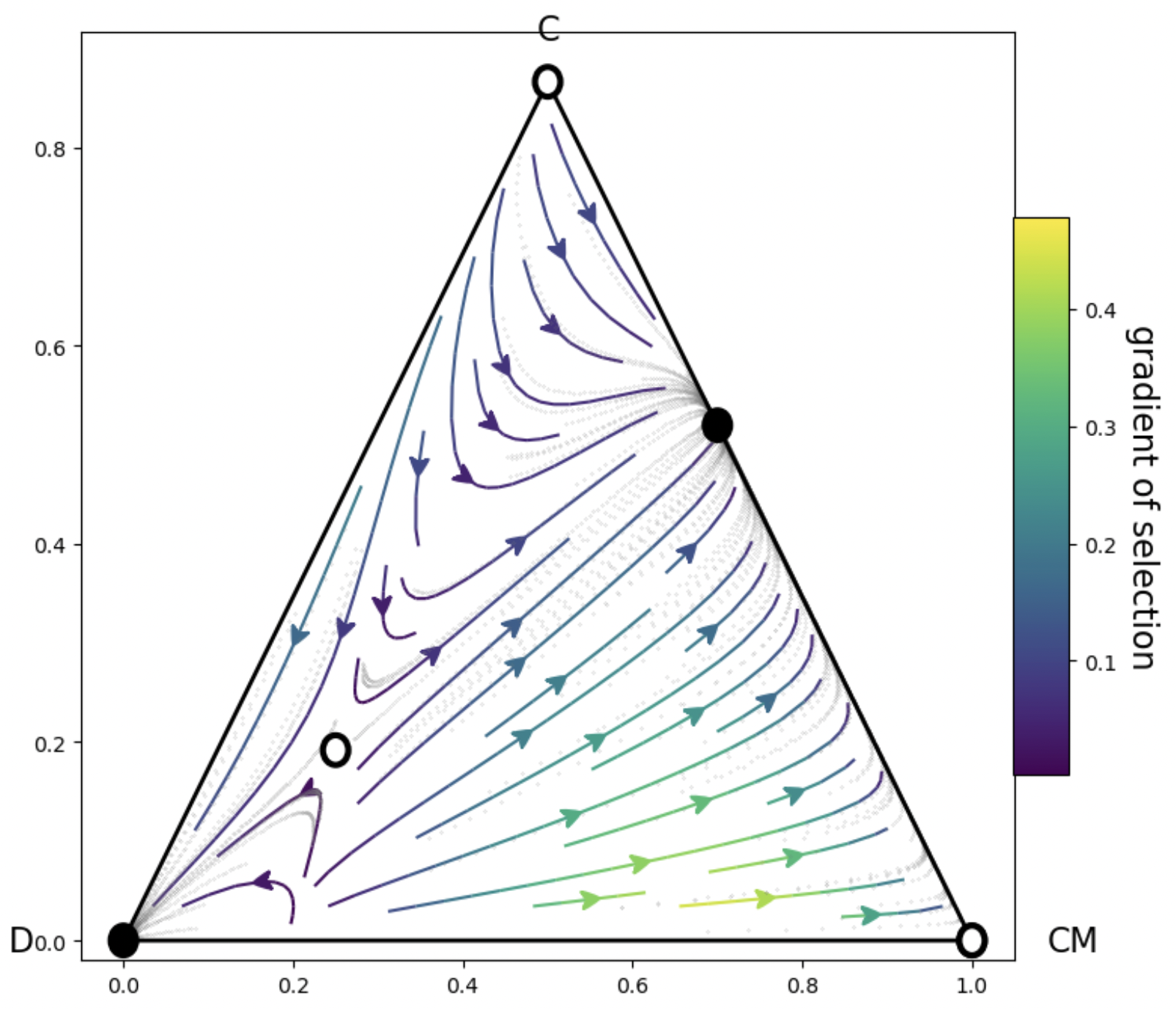}} 
    \subfigure[c]{\includegraphics[width=0.29\textwidth]{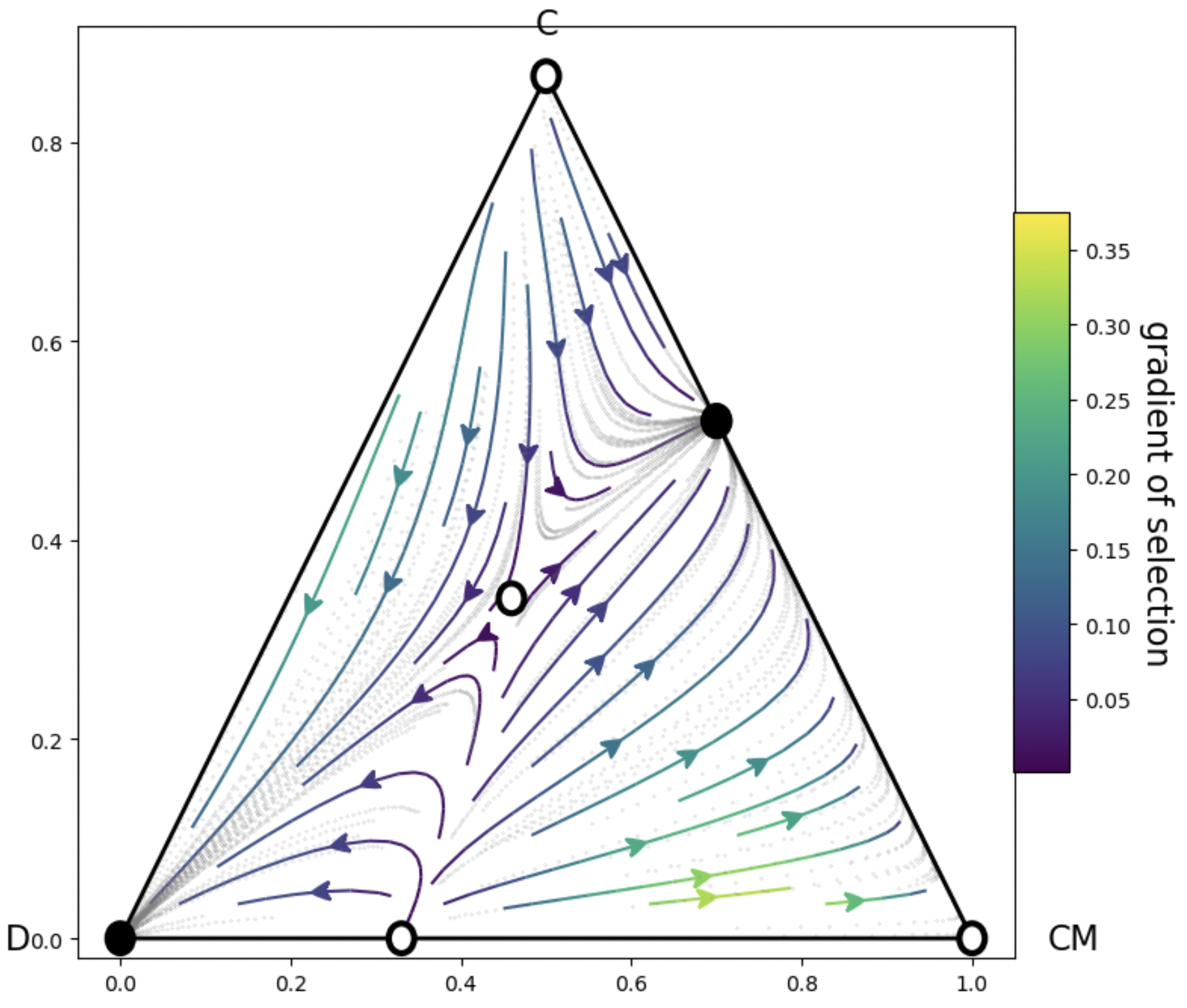}}
    \caption{(a) Control (b) Agents obey an inverted S shape distortion in how they estimate expected punishment (overestimate small probabilities and underestimate large ones) (c) Agents obey an S shape distortion in how they estimate expected punishment (underestimate small probabilities and overestimate large ones)}
    \label{fig3}
\end{figure}

Under the inverted S distortion, similar to the coarse overestimation bias, we see an increase in the size of the attractor for cooperation, which therefore makes cooperation more likely. When we have an inverted S distortion and a decrease in the size of the cooperative attractor when it is S shaped, which makes cooperation more unlikely (Fig.\ref{fig3}). This is interesting as it suggests incorporating these findings from psychophysics into game theory changes collective outcomes.

However, despite its more detailed empirical grounding, the result here is hard to interpret as it hinges on which curve (S or inverse S) humans use when forming institutions. The empirical evidence for this is not clear-cut. For example, it seems that the use of S or inverse S-shaped curves is heterogeneous in population, so not everyone has the same shape \cite{zhang_bounded_2020}. Secondly, it also depends on the type of task, i.e. motor decision task vs abstract economic decision task \cite{zhang_bounded_2020}. Furthermore, social contexts can change outcomes, e.g. when people are told they are playing a game with people vs a computer, their behaviour changes \cite{rilling_neural_2004} and whether a participant can see if they are playing against a real person or not also affects behaviour \cite{duguid_coordination_2014}. Such experiments on probability distortion have not been attempted in a truly social setting. Therefore, understanding the relation between probability distortion and collective outcome is not straightforward.

\subsection{Noise (bounded uncertainty)}
Given the difficulty of interpretation of empirical results for our institution's bootstrapping problem. We now attempt to motivate this in an alternate fashion using the inherent uncertainty brought about because individuals are bounded by their local context.
It has been established that individual agents are bounded \cite{simon_behavioral_1955} and often limited to the information around them. Therefore, they often anchor on local estimates and recent experiences, e.g. people who know someone with cancer have higher estimates of its prevalence \cite{simon_behavioral_1955,pachur_how_2012}. These estimates are subject to noise for a host of different factors, e.g. different life experiences, affective makeups of each individual, social networks, social media bubbles, etc. \cite{sundh_human_2024,kahneman_noise_2021,pachur_how_2012}. 

In our model, an agent may be very optimistic with respect to being punished based on its life experience, and may have been lucky enough not to have been punished before and be in a social group that either doesn't get punished very often or doesn't like advertising it to others. Alternatively, another agent could have been punished quite a few times and be in a social group that gossips about punishment often, leaving it with a rather bleak estimate of how likely it is to get punished (overestimates risk). Because of the uncertainty inherent in being bounded, agents have to anchor on local cues, which results in a spectrum of differing estimates of being punished across the population.

To model the inherent uncertainty of individuals bounded by different contexts and noisy perception, we represent it as a uniform distribution of biases (away from the perfect estimate) across the population. Note that this is not the same as all individuals having the same bias as in previous sections; here, each individual has a different bias determined by their local social and physical context. Further, there is no skew in the bias at the population level; all noisy estimates even out to be an accurate, unbiased group estimate, as supported by collective intelligence studies \cite{galton_vox_1907,krause_swarm_2011,kao_counteracting_2018}. 

To model this, we need a stochastic payoff for the defect strategy. To do this, we sample uniformly in a range with 40 samples (equal to the population) centred at one. Which is then multiplied by $C_f$ so that effectively each agent has their own estimate of the expected cost of freeriding, with equal chances of them under- or overvaluing the probability of being punished. 

\begin{figure}[hbt!]
    \centering
    \subfigure[a]{\includegraphics[width=0.29\textwidth]{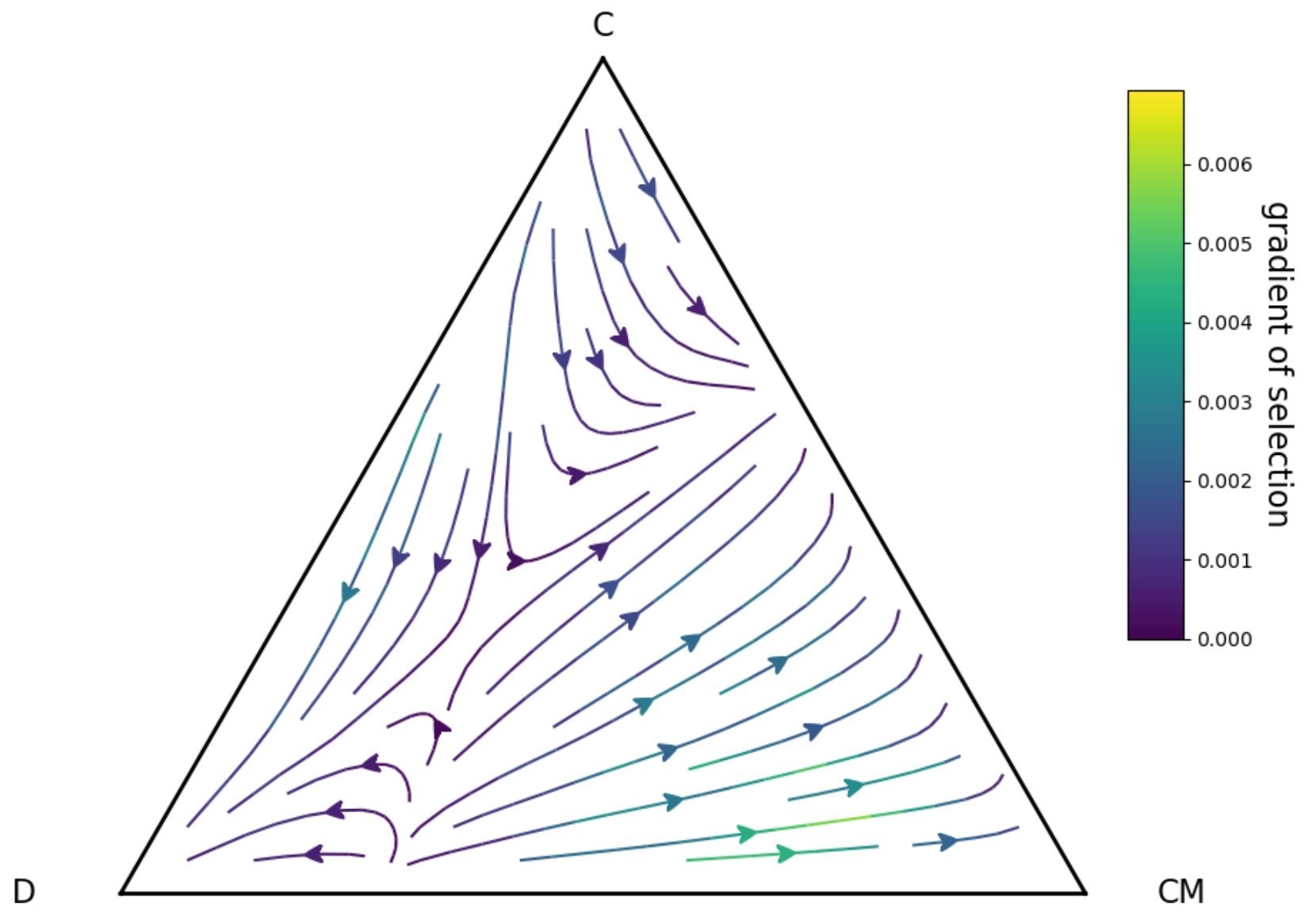}} 
    \subfigure[b]{\includegraphics[width=0.29\textwidth]{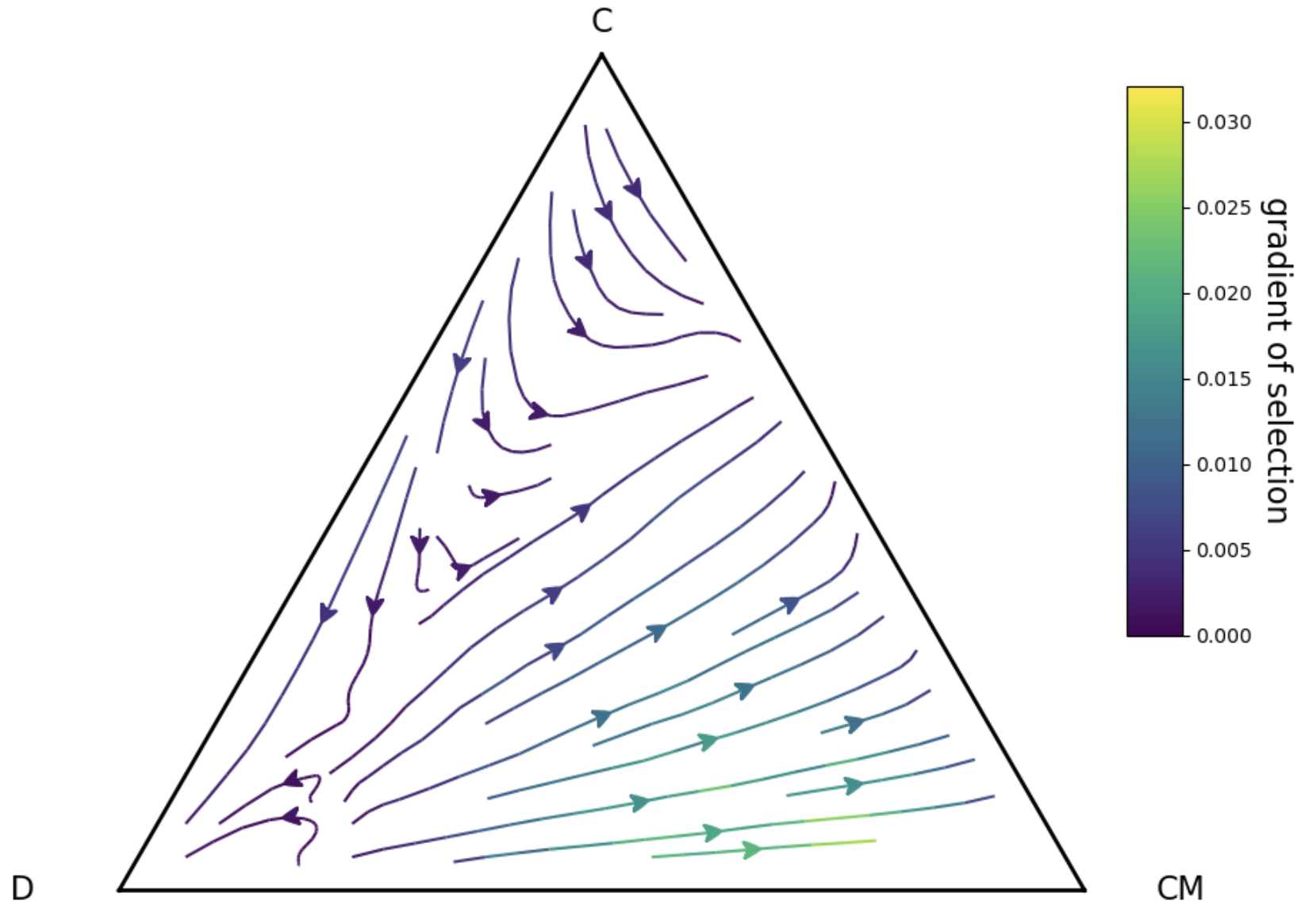}} 
    \subfigure[c]{\includegraphics[width=0.29\textwidth]{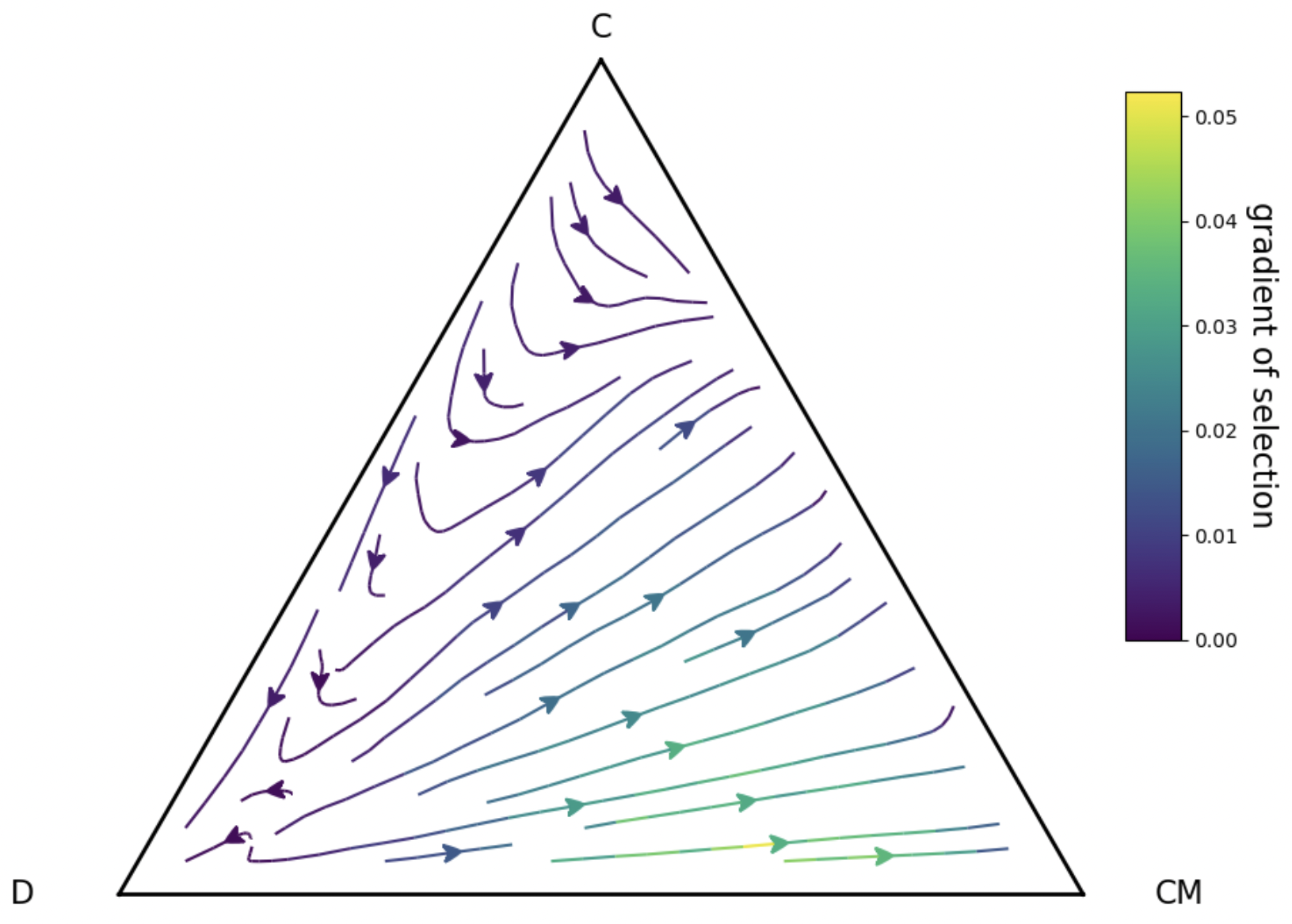}}
    \caption{(a) Control (b) noisy perception in range [0.25:4] centred at 1 (c) noisy perception in range [0.125:8] centred at 1. Note that due to the stochastic nature of the Moran process used to model these dynamics, we cannot derive stable and unstable attractors (denoted by black and white dots) as we could with the deterministic replicator equation used in figures 1-3.}
    \label{fig4}
\end{figure}

We see that, surprisingly, under this noise, cooperative attractors are larger (Fig.\ref{fig4}). This doesn’t solve the bootstrapping problem completely, but it means stochastic extrinsic shocks, e.g. "strong leader" mutation or an influx of cooperators, are more likely to trigger an institution to be formed. This is striking because the system is not biased in any particular direction and despite this still favours cooperation, i.e., the institution being formed. This means that solely taking into account the fact of being a bounded agent in a noisy world increases the chances of an institution being formed. 

The effect is more extreme for a larger range of noise, suggesting that a more diverse range of perceptions, due to the numerous factors affecting perception, further increases cooperation (Fig.\ref{fig4}c).

\subsubsection{Why does unbiased noise lead to a biased outcome?}

Despite the noise's unbiased nature, there is an asymmetry in the system. This is due to the proportional nature of the bias. To create a bias for each agent, we either make them underestimate the expected cost of being punished by taking the product of a value less than 1, or make them overestimate it by taking it to above 1. Underestimating bias when the number of monitors is near zero does not change the estimate much, but overestimating has a larger net effect. Thus overpowering the effect of underestimation. This effect does not exist on the other side of the range (when there are many monitors) since there isn't a limit on how high one can overestimate.

\subsection{Absolute noise (bounded uncertainty)}

In the above section, we described the effect of a noisy proportional bias where the utility of an action is either over or under-valued in a proportional manner, but we could also implement this as absolute noisy perception, where an absolute value is added or subtracted to the utility of an action to produce a noisy expectation i.e. it can be in a range -8:8, -16:+16 which is added to the $Cf$ estimate. 

\begin{figure}[hbt!]
    \centering
    \subfigure[a]{\includegraphics[width=0.29\textwidth]{moran_control.jpeg}} 
    \subfigure[b]{\includegraphics[width=0.29\textwidth]{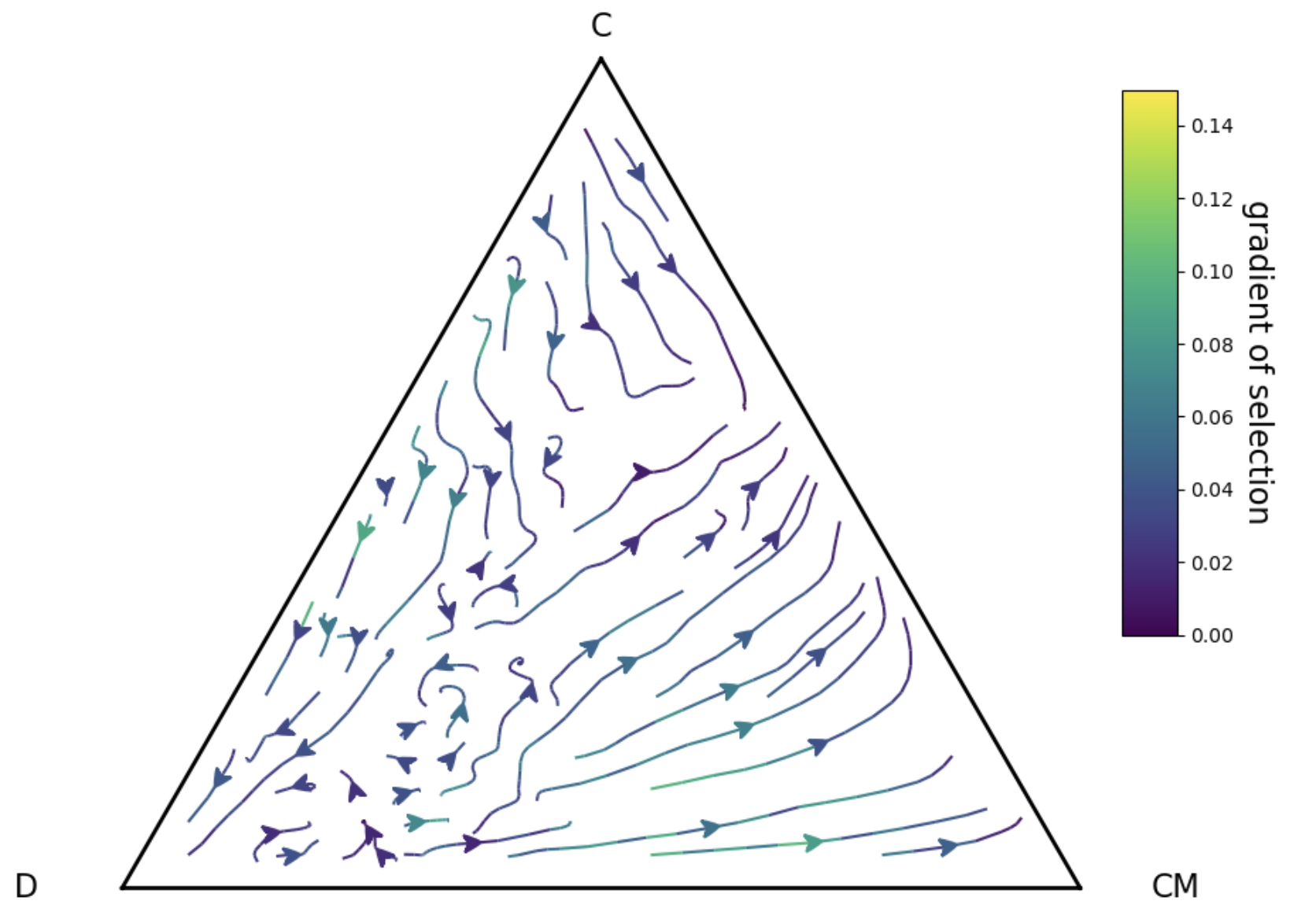}} 
    \subfigure[c]{\includegraphics[width=0.29\textwidth]{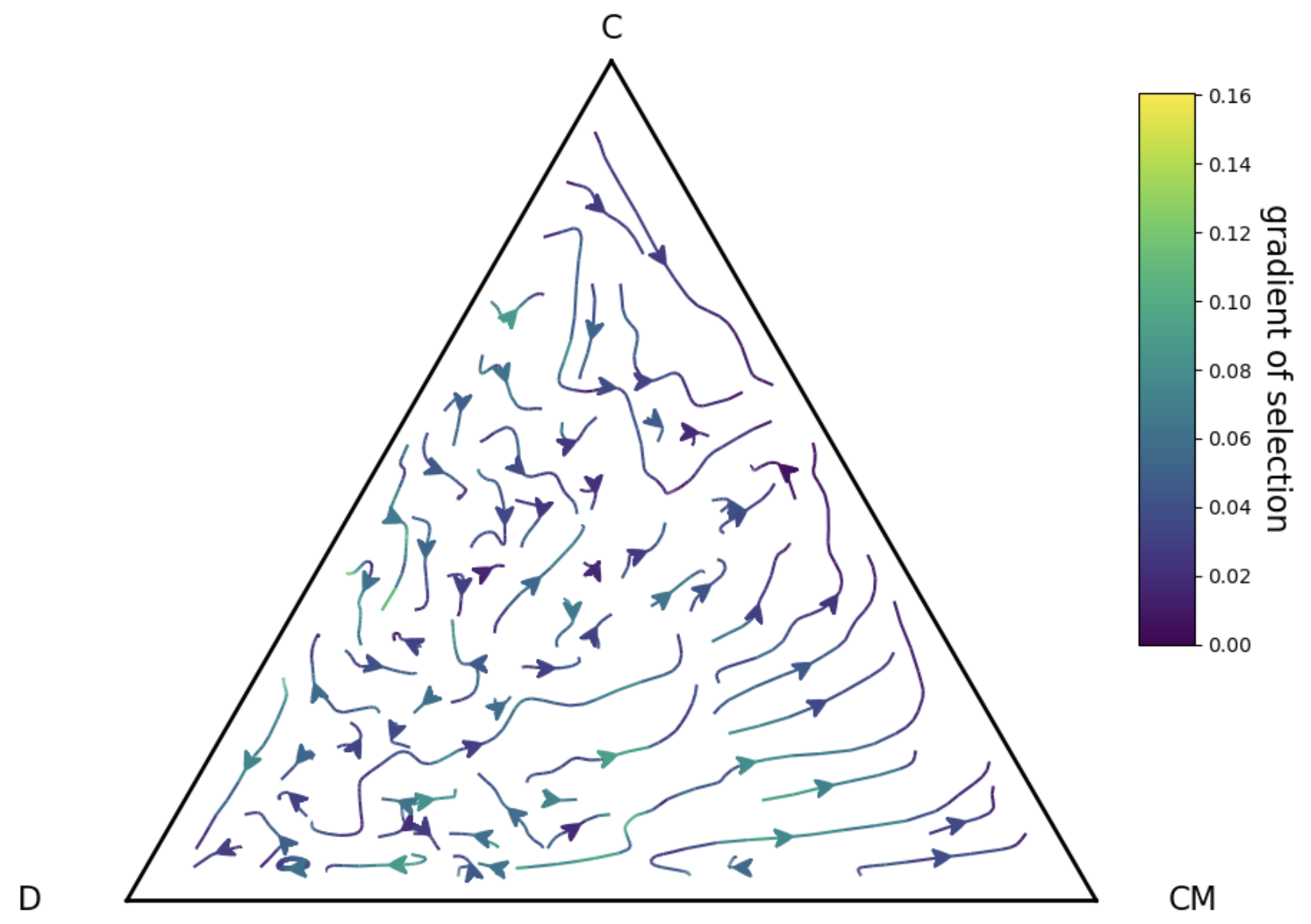}}
    \caption{(a) Control (b) absolute noisy perception range [-8:8] (c) absolute noisy perception range [-16:16]). Note that due to the stochastic nature of the Moran process used to model these dynamics, we cannot derive stable and unstable attractors (denoted by black and white dots) as we could with the deterministic  replicator equation used in figures 1-3.}
    \label{fig5}
\end{figure}

Under absolute noisy bias, we get a qualitatively different effect emerging (Fig.\ref{fig5}), instead, it seems that the shifts in strategy space just become more uncertain. We see that regions are more mixed, and certain regions 
in strategy space that are far away from each other are more likely to be linked. This kind of noisy bias can help solve the bootstrapping problem since it can result in sudden shifts from non-cooperative to cooperative equilibria. However, there would need to be a mechanism to reduce noise in estimates once a cooperative state is reached, to prevent the institution from falling apart due to noise.

\section{Discussion}

\subsection{Summary}

Even when an institution is favourable to the individuals who constitute it, there is still the issue of reaching the critical mass of people who join it to incentivise its creation: no one wants to join and then be let down by no one else joining. We call this the institution bootstrapping problem. 

To overcome this problem, we ask, what if the perceptual make-up of individuals makes them more likely to join an institution? And can merely believing an institution exists be enough to create it?
Incorporating the psychological literature of subjective probability estimation and cognitive biases, we relax the assumption of perfect perception in game theory to investigate this.
We show that a coarse bias in the subjective probability of being punished can make an institution either more likely or unlikely to form, depending on whether agents have a bias that makes them either overestimate or underestimate the subjective probability of being punished. 

Further, we incorporate empirical work from psychophysics \cite{zhang_bounded_2020} which shows that individuals distort probabilities at the extremes of the range, biasing subjective probability of events when they are either highly unlikely or highly likely. We incorporate this and show that it also influences the threshold of monitors needed to form an institution in a similar fashion to a coarse-grained bias. However, as described in the results, empirical work is unclear concerning how these biases work in social settings \cite{zhang_bounded_2020,rilling_neural_2004}.

We also find that, surprisingly, having a noisy perception across the population (no bias at the group level), which is equivalent to individuals being biased by their bounded position in the social and physical world, lowers the threshold of individuals needed to start an institution. Therefore, this bounded noise is beneficial for forming a cooperative institution. Just the bare bone limitations of being a bounded agent embedded in a social world (hence noise) increase the chance of institutions being formed and make them more robust.
This is if the noisy perception is proportional; however, if it is absolute, it generally makes the system more likely to shift states in an unpredictable manner (either to form an institution or not). This can be useful; however, there would need to be a way to eliminate this noisy absolute bias once the system is in a cooperative state, i.e. agents would need to agree on a ground truth when conditions are favourable for cooperation.

Moreover, by relaxing game theory's assumptions, which describes phenomena with group level State Variable Models (SVM) \cite{smith_evolution_1982}, e.g., the average expected cost of free riding. We show that even if you have a noisy uniform perception that cancels out and is therefore equivalent to the game-theoretic SVM, the variance of perceptions can still change the outcome of the analysis.

\subsection{Implications of this finding}

Our findings suggest that: making agents have a more accurate estimate of the world, although it seems an intuitively good idea, may make them less likely to form an institution. Uncertainty and boundedness are therefore important to have. This aligns with findings in other contexts, which also suggest that imperfect information may be better for collective outcomes \cite{salge_dont_2014,vogrin_confirmation_2023,davies_if_2011}.

Further, we assume in this paper that the institution is beneficial to all agents who join it. But there may also be cases where agents are fooled into joining an institution that benefits some but hurts others, e.g., a corrupt institution where monitors take the collective resource for themselves.
In this case, how would agents disobey an unjust institution \cite{burth_kurka_disobedience_2018}? In these cases, you may want perfect/better information and for agents to agree with one another in order to rise up against the institution. This could be done if bounded agents come together to combine their estimates and take the mean/median, as done in collective intelligence studies \cite{kao_counteracting_2018}, and then, if they all use this group estimate in their decision-making, overcome a corrupt institution. Another approach is to consider other biases that may enable the collective action needed to topple an unjust institution. For example, groupthink is often regarded as defective. However, due to the homogenisation of opinions \cite{turner_twenty-five_1998}, it may be beneficial, as it would reduce variance in estimates, perhaps allowing agents to withdraw from the unjust institution.


Furthermore, we can connect our model to collective action models \cite{gintis_individuality_2016}, e.g. voting, taking political action, or protesting. If we interpret the cost of contributing as the cost that may be involved in collective action, e.g. group effort, personal sacrifice, etc., then noise (bounded uncertainty) makes the institution (collective action) more likely, and having too much information (or too accurate an estimate) makes things hopeless, and agents stop trying to join the collective action. This corroborates well with Herbert Gintis' argument \cite{gintis_individuality_2016} that rational actors with perfect information are less likely to act since they perceive their actions will make no difference in large groups. He then asks: Why do we see collective action even when it is irrational? Gintis then argues, drawing from the anthropological literature, that humans have an evolved bounded psychology that is tuned for small groups, which changes their estimates of making a difference at the collective level and, therefore, explains the prevalence of collective action in human societies.


In conclusion, as shown in this paper, agents' cognitive limitations may be a benefit rather than a hindrance. What looks irrational for the individual may be beneficial for the group, and ultimately, for the individual (groups where individuals support each other are often more successful than individuals on their own).

Our general model can thus serve as an initial template to assess the impact of cognitive biases and uncertainty in more specific cases of institutions in socio-technical systems, e.g. shared online servers, data sharing systems, and neighbourhood power cooperatives \cite{savarimuthu_norm_2011,nardin_classifying_2016,morris-martin_norm_2019}.

\subsection{Our results contradict the conventional wisdom that noise is detrimental to cooperation
}

Early influential game theoretic and agent-based models showed noise is deleterious for cooperation and leads to a breakdown of the tit-for-tat mechanism, as it leads to rounds of mutual defection \cite{bendor_when_1991,axelrod_further_1988}. 

Further, in an empirical study show that Salahshour et al \cite{salahshour_cost_2022} show that stochastic punishment (in this case the punishment factor - how much the fine is multiplied, which would be equivalent to having a noisy $s$ in our model) reduces group contributions to the game and encourages more antisocial punishment (punishment of cooperators instead of defectors, which in turn disincentivises cooperation).

This discordance in our current paper about the positive collective benefits of noise may be explained by the fact that, in our model, it is not noise in the punishment but noise in the perception of the expected punishment that engenders cooperation, an important difference. These contextual factors make it hard to make general statements on the effects of noise on cooperation, one way or another. The nuance of uncertainty and noise and its non-trivial impact on norms and institutional emergence should be considered in the design of normative multi-agent systems \cite{savarimuthu_norm_2011,nardin_classifying_2016,morris-martin_norm_2019}. 

\subsection{Limitations and extensions}
So far, we only have one expected value in our model (expected punishment).
However, there are other world variables that agents could have biased or noisy perceptions of, e.g. cost of monitoring/cost of contribution. This can be altered in an agent-based model, where it is possible to dissociate perceived and real payoff as in \cite{mcnamara_learning_2021}. This allows the exploration of what factors in the population, e.g. scarcity of resources, lead to which biases/noise to evolve, e.g. loss aversion bias \cite{mcnamara_risk-sensitive_1992,anagnou_effect_2023}

Further, although using EGT is a rigorous way to study the value of choices in a social environment and hence the direction in strategy space the system will move in. It stays agnostic to complex behaviours and interactions between individual and population-level dynamics. Agent-based models have the ability to capture complex behaviours and interactions in
executable form, and to explore emergent phenomena simply by "running” variants of the model \cite{epstein_growing_1996}. 

For example, a candidate mechanism for why uncertainty may encourage the joining of institutions: an initial proportion of individuals with noisy perception will be biased towards overestimating the probability of being caught, they will change their behaviour, which will in turn cause others who are less biased to do the same since there will now be more members in the institution, this process repeats instantiating a feedback loop that allows agents to bootstrap to cooperation.

\begin{credits}
\subsubsection{\ackname} Stavros Anagnou was supported by a PhD studentship from the University of Hertfordshire and a MITACS Globalink Research Award grant. This research was made possible, in part, thanks to the Canada Research Chairs program. We would also like to thank the anonymous reviewers for their feedback on the manuscript and Elias Fernández Domingos for advice regarding the EGTtools library.

\subsubsection{\discintname}
The authors have no competing interests to declare that are relevant to the content of this article.
\end{credits}
%
%
%
\bibliographystyle{splncs04}
\bibliography{references_institutions}

\end{document}